\documentclass{elsart}
\journal{} 

\usepackage{epsfig}
\usepackage{amssymb}

\begin{document}
\begin{frontmatter}

\title{Pulse shapes from electron and photon induced events in 
segmented high-purity germanium detectors}
\author{I.~Abt}, 
\author{A.~Caldwell}, 
\author{K.~Kr\"oninger\corauthref{cor}}\ead{kroening@mppmu.mpg.de}, 
\author{J.~Liu}, 
\author{X.~Liu}, 
\author{B.~Majorovits} 
\address{Max-Planck-Institut f\"ur Physik, M\"unchen, Germany}
\corauth[cor]{Max-Planck-Institut f\"ur Physik, M\"unchen, Germany, 
Tel. +49-(0)89-32354-337}

\begin{abstract} 
Experiments built to search for neutrinoless double beta-decay are
limited in their sensitivity not only by the exposure but also by the
amount of background encountered. Radioactive isotopes in the
surrounding of the detectors which emit gamma-radiation are expected
to be a significant source of background in the GERmanium Detector
Array, {\sc GERDA}. \\

Methods to select electron induced events and discriminate against
photon induced events inside a germanium detector are presented in
this paper. The methods are based on the analysis of the time
structure of the detector response. Data were taken with a segmented
{\sc GERDA} prototype detector. It is shown that the analysis of the
time response of the detector can be used to distinguish multiply
scattered photons from electrons.
\end{abstract} 
\begin{keyword}
double beta-decay, germanium detectors, segmentation, pulse shape analysis 
\PACS 23.40.-s \sep 14.60.Pq \sep 29.40.-n
\end{keyword}
\end{frontmatter} 


\section{Introduction}
\label{section:introduction}

\enlargethispage{0.5 cm} 

Radioactive decays in which photons with energies above $Q=2039$~keV
are emitted are expected to be a significant source of background for
the GERmanium Detector Array, {\sc GERDA}~\cite{proposal}. {\sc GERDA}
is an experiment which is currently being constructed and has as aim
the search for the neutrinoless double beta-decay ($0\nu\beta\beta$)
of the germanium isotope $^{76}$Ge. \\ 

Methods to distinguish between electrons and multiply scattered
photons using the time structure of the germanium detector response,
or pulse shape, are presented in this paper. The pulse shape depends
on the location and the spatial distribution over which energy is
deposited inside the detector in a single event. Photons in the
MeV-energy region will predominantly Compton-scatter and deposit
energy at locations separated by centimeters. These events are
referred to as {\it multi-site events}. In contrast, electrons in the
same energy region have a range of the order of a millimeter. Events
of this kind are referred to as {\it single-site events}. \\

Pulse shape analysis methods have been developed for nuclear
experiments such as {\sc AGATA}~\cite{agata} and {\sc
GRETA}~\cite{greta} as well as for double beta-decay
experiments~\cite{Hellmig:2000xp,HM,IGEX,Aalseth:2000hy,Elliott:2005at}. In
the context of the latter these techniques are now extended to
segmented detectors. In this study the focus is on the pulse shape
analysis after the application of a single segment requirement as
presented in~\cite{Abt:2007rg}. The performance of the pulse shape
analysis with and without segment information is compared based on
data taken with an 18-fold segmented {\sc GERDA} prototype
detector. \\

The experimental setup and the collected data sets are described in
Section~\ref{section:setup}. The accompanying Monte Carlo simulation
is introduced in Section~\ref{section:simulation}. A parameter
accessible in simulations which is a measure of the volume over which
energy is deposited inside the detector is defined. A definition of
single-site and multi-site events is derived from the Monte Carlo data
sets based on this parameter.  The fraction of single-site and
multi-site events in the data sets is estimated. Three analysis
methods are presented in Section~\ref{section:methods} and these
methods are applied to the data sets taken with the prototype
detector. The results are summarized in
Section~\ref{section:results}. Conclusions are drawn in
Section~\ref{section:conclusions}.


\section{Experimental setup and data sets} 
\label{section:setup}

\subsection{Experimental setup and data taking} 

The segmented germanium detector under study is the first segmented
{\sc GERDA} prototype detector. The true coaxial cylindrical crystal
has a height of 70~mm, an outer diameter of 70~mm and a central bore
with a diameter of 10~mm. It is 18-fold segmented with a 6-fold
segmentation in the azimuthal angle $\phi$ and a 3-fold segmentation
in the height $z$. It was operated in a conventional test
cryostat. Signals from the core and the segment electrodes were
amplified and subsequently digitized using a 14-bit ADC with a
sampling rate of 75~MHz. The energy and the pulse shapes of the core
and the 18~segment electrodes were recorded for each event. The pulse
shape data consists of 300 13.3~ns samples of the integrated charge
amplitude. The onset of the signal was delayed by one~$\mu$s. The
(full width at half maximum) energy resolution of the core electrode
was 2.6~keV at energies around 1.3~MeV, the energy resolutions of the
segment electrodes ranged from 2.4~keV to 4.8~keV with an average
segment energy resolution of 3.3~keV. Details of the experimental
setup and the detector performance can be found
in~\cite{Abt:2007rf}. \\
 
\begin{figure}[ht!]
\center
\mbox{\epsfig{file=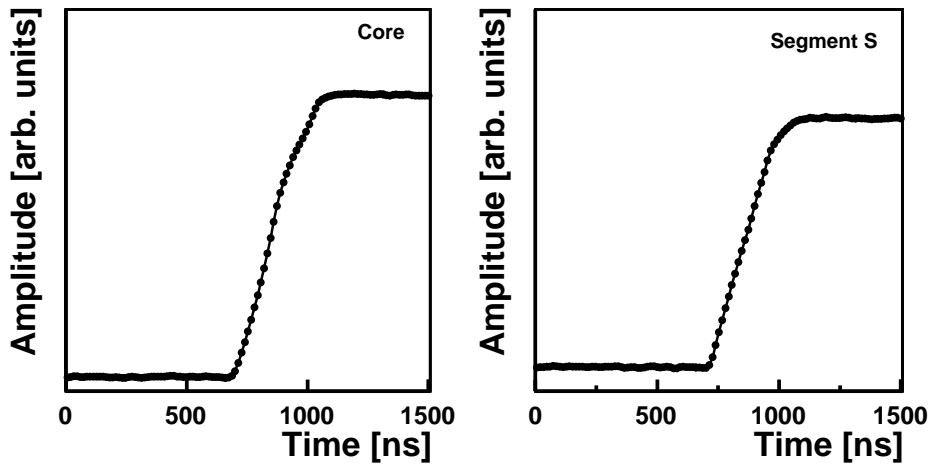,width=0.9\textwidth}}
\caption{Pulse shape measured with the core (left) and with the
segment~$S$ electrodes (right) for an event in the segment data
set. The energy of $1758$~keV seen in the core is completely contained
in segment~$S$. The starting time is chosen arbitrarily in this
example. The amplitude is in arbitrary units but the scale is the same
for both pulse shapes. The pulse shapes are dominated by different
charge carrier types.
\label{fig:example}}
\end{figure}

A 100~kBq $^{228}$Th source was placed at $z=0$~cm and $r=17.5$~cm
with respect to the detector center ($z=0$~cm, $r=0$~cm) facing
towards the center of a segment, $S$, located in the middle row. Two
data sets were taken with different trigger conditions labeled
$TR_{C}$ and $TR_{S}$. The former trigger condition requires the core
electrode to show an energy above 1~MeV. The collected data set is
referred to as {\it core data set} and contains $127\,000$~events. The
latter trigger condition requires segment~$S$ to show an energy above
1~MeV. The collected data set is referred to as {\it segment data set}
and contains $420\,000$~events. As an example,
Figure~\ref{fig:example} shows a pulse shape measured with the core
(left) and with the segment~$S$ electrode (right) for an event in the
segment data set. The core-energy spectra will be shown in
Section~\ref{subsection:application}. \\

\pagebreak 

\subsection{Event selection} 
\label{subsection:selection} 

A pre-selection applied to the segment data set collects events with
energy deposited only in one segment. It requires the energy measured
in segment~$S$ to be the same as the energy measured in the core
within $\pm5$~keV, according to about $\pm 4~\sigma$ given the energy
resolution. In total, $150\,396$~events fulfill the pre-selection
criterion. \\

Four data samples each are selected from the core and segment data
sets. The data samples are defined by the energy measured in the core
and are labeled:
\begin{itemize}
\item $DEP$: The sample contains events with a core energy in the region 
of \mbox{$(1593\pm5)$~keV}. These events are associated with the
double escape peak of the $2615$~keV $^{208}$Tl photon. The photon
produces electron-positron pairs of which the positron subsequently
annihilates. Both 511~keV annihilation photons escape the
detector. The energy is predominantly deposited on a millimeter-scale;
i.e., locally.
\item $\Gamma_{1}$: The sample contains events with a core energy in the 
region of \linebreak \mbox{$(1620\pm5)$~keV}. These events are
associated with photons of this energy produced in the decay of
$^{212}$Bi. The photons mostly scatter multiple times before their
energy is fully deposited inside the detector.
\item $\Gamma_{2}$: The sample contains events with a core energy in the 
region of \linebreak \mbox{$(2615\pm5)$~keV}. These events are
associated with photons of this energy produced in the decay of
$^{208}$Tl. The photons mostly scatter multiple times before their
energy is fully deposited inside the detector.
\item $ROI$: The sample contains events with a core energy in the region 
of interest, $(2039\pm50)$~keV. These events are predominantly
associated with Compton-scattered photons from $^{208}$Tl.
\end{itemize}

The requirements of the trigger, pre-selection and event selection are
listed in Table~\ref{table:datasets}. Also the number of events in the
corresponding data samples are shown. The amount of background in each
data sample, as estimated from taking spectra without the $^{228}$Th
source present, was found to be less than 1\%. \\

\begin{table}[ht!]
\caption{Requirements of the trigger, pre-selection and event selection, 
and the number of events in the corresponding data samples. $E_{C}$
and $E_{S}$ are the energies seen in the core and in segment~$S$,
respectively.
\label{table:datasets}}
\center
\begin{tabular}{llr} 
\\ 
\hline
Cut                      & Condition                                                          & Events \\ 
\hline
Trigger ($TR_{C}$)       & $E_{C} > 1$~MeV                                                 & $127\,000$ \\ 
Pre-selection            & -                                                               & $127\,000$ \\ 
Selection ($DEP$)        & $\left| E_{C} - 1593\mathrm{~keV} \right| < \phantom{0}5$~keV & $1673$ \\ 
Selection ($\Gamma_{1}$) & $\left| E_{C} - 1620\mathrm{~keV} \right| < \phantom{0}5$~keV & $1965$ \\ 
Selection ($\Gamma_{2}$) & $\left| E_{C} - 2615\mathrm{~keV} \right| < \phantom{0}5$~keV & $22\,924$ \\ 
Selection ($ROI$)        & $\left| E_{C} - 2039\mathrm{~keV} \right| < 50$~keV           & $6\,431$ \\ 
\hline
Trigger ($TR_{S}$)       & $E_{S} > 1$~MeV                                                 & $420\,000$ \\ 
Pre-selection            & $\left| E_{C} - E_{S} \right| < 5$~keV                          & $150\,396$ \\ 
Selection ($DEP$)        & $\left| E_{C} - 1593\mathrm{~keV} \right| < \phantom{0}5$~keV & $3492$ \\ 
Selection ($\Gamma_{1}$) & $\left| E_{C} - 1620\mathrm{~keV} \right| < \phantom{0}5$~keV & $1972$ \\ 
Selection ($\Gamma_{2}$) & $\left| E_{C} - 2615\mathrm{~keV} \right| < \phantom{0}5$~keV & $19\,243$ \\ 
Selection ($ROI$)        & $\left| E_{C} - 2039\mathrm{~keV} \right| < 50$~keV           & $7707$ \\ 
\hline
\end{tabular} 
\end{table} 


\section{Monte Carlo simulation} 
\label{section:simulation} 

The GEANT4~\cite{geant4} based MaGe~\cite{MaGe} framework was used to
simulate the prototype detector setup (for details and a validation of
this particular simulation see~\cite{Abt:2007rg}). A Monte Carlo study
was performed to estimate the spatial distribution over which energy
is deposited in the detector for events in the different data
samples. A $^{228}$Th source was simulated. The trigger, pre-selection
and event selection requirements discussed in the previous section
were applied to the Monte Carlo data. The data sets are referred to as
{\it core} and {\it segment Monte Carlo data sets}.

\begin{figure}[ht!]
\center
\begin{tabular}{cc}
\begin{minipage}[ht!]{0.45\textwidth}
\mbox{\epsfig{file=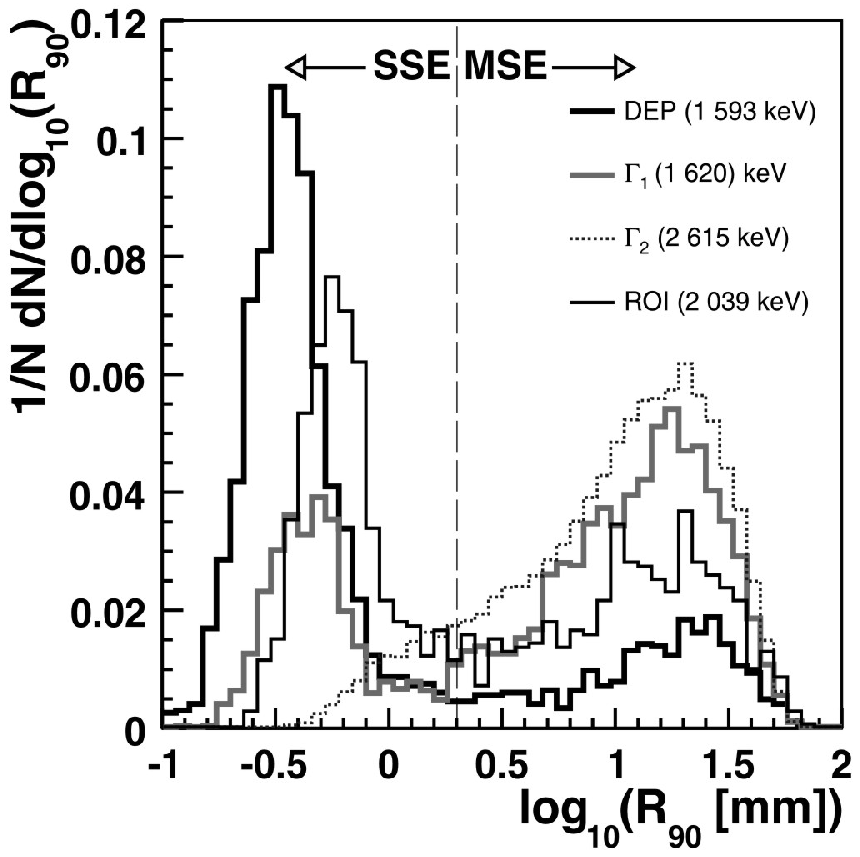,width=\textwidth}}
\end{minipage}
&
\begin{minipage}[ht!]{0.45\textwidth}
\mbox{\epsfig{file=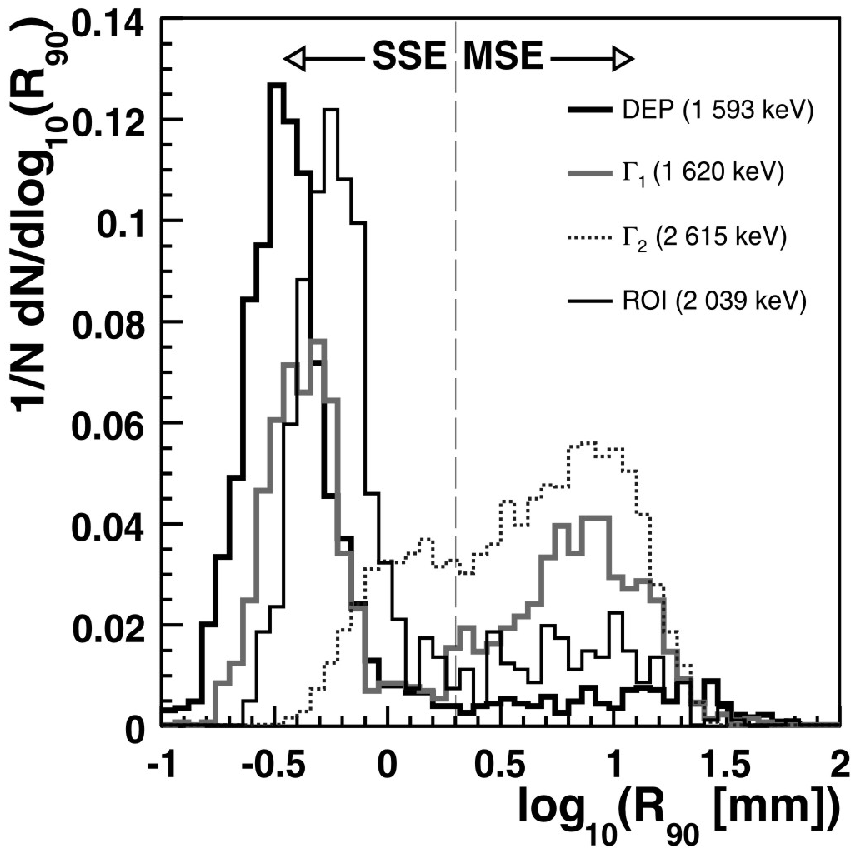,width=\textwidth}}
\end{minipage} \\
\end{tabular}
\caption{Normalized distributions of $R_{90}$ for the $DEP$, 
$\Gamma_{1}$, $\Gamma_{2}$ and $ROI$ samples for the Monte Carlo core
(left) and segment data sets (right). Single-site events (SSE) and
multi-site events (MSE) are defined by requiring $R_{90} < 2$~mm and
$R_{90} > 2$~mm (dashed line) as discussed in the text.
\label{fig:R90}}
\end{figure}

A measure for the spatial distribution over which energy is
distributed inside the detector is the radius $R_{90}$. This is
defined as the radius inside which 90\% of the energy in a single
event is deposited; for a detailed discussion
see~\cite{segmentation}. Figure~\ref{fig:R90} shows the distribution
of $R_{90}$ for the $DEP$, $\Gamma_{1}$, $\Gamma_{2}$ and $ROI$
samples for the core (left) and segment (right) Monte Carlo data
sets. All distributions are normalized to unity. The $R_{90}$
distributions range from 0.1~mm ($\log_{10}(R_{90})=-1$) up to 7~cm
($\log_{10}(R_{90})=1.8$). The $DEP$ samples are dominated by events
with $R_{90}$ in a region from 0.1~mm to 1~mm. A long tail towards
larger radii is visible and mostly due to events in the underlying
Compton-shoulder of $^{208}$Tl and events in which electrons undergo
hard bremsstrahlung processes. The $R_{90}$ distributions for the
$\Gamma_{1}$ and $ROI$ samples have two prominent regions each, one at
radii from 0.3~mm to 1~mm and a second from 3~mm to 6~cm.  The latter
one is due to multiply scattered photons whereas the former is due to
photons with higher energy which only scatter once and then leave the
detector. The $R_{90}$ distributions for the $\Gamma_{2}$ samples
range from 0.3~mm to about 7~cm with a maximum at around 2~cm for the
core Monte Carlo data sample and at around 1~cm for the segment Monte
Carlo data sample. The sample is dominated by events in which photons
scatter multiple times. No peak at small $R_{90}$ is visible. \\

It is expected that the single segment requirement rejects events with
large values of $R_{90}$. Indeed, the distributions of $R_{90}$ in the
segment Monte Carlo data samples are suppressed in the region above
1~cm. The peaks between 0.1~mm and 1~mm in the $DEP$, $\Gamma_{1}$ and
$ROI$ samples are more pronounced in this case. \\

Single-site and multi-site events are defined by requiring
$R_{90}<\overline{R}$ and $R_{90}>\overline{R}$, respectively, where
$\overline{R}$ is a chosen parameter value. The distributions of
$R_{90}$ for the $DEP$ samples suggest $\overline{R}=2$~mm
($\log_{10}(\overline{R})=0.3$). Also, due to the sampling rate of
75~MHz and the average drift velocity of charge carriers
($\mathcal{O}(10^{8})$~mm/s) energy deposits closer than about 2~mm
cannot be resolved. The fractions of single-site events in the Monte
Carlo data samples are thus defined and summarized in
Table~\ref{table:fraction}. Also listed are the corresponding
systematic uncertainties of the fractions which are derived by varying
the parameter $\overline{R}$ by $\pm1$~mm.

\begin{table}[ht!]
\caption{Fractions of single-site events in the Monte Carlo data samples. 
The errors are derived by varying the parameter $\overline{R}$ by
$\pm1$~mm.
\label{table:fraction}}
\center
\begin{tabular}{lcccc}
\\
\hline
Monte Carlo data samples & $DEP$ & $\Gamma_{1}$ & $\Gamma_{2}$ & $ROI$ \\
& ($1593$~keV) & ($1620$~keV) & ($2615$~keV) & (2039~keV) \\
\hline 
Core samples    & $(77.9^{+1.6}_{-3.4})$\% & $(30.5^{+4.0}_{-3.6})$\% & $(12.2^{+\phantom{0}6.0}_{-\phantom{0}7.6})$\%  & $(52.4^{+3.8}_{-7.6})$\% \\ 
Segment samples & $(89.0^{+1.1}_{-3.0})$\% & $(55.0^{+5.0}_{-4.4})$\%  & $(30.0^{+10.0}_{-16.8})$\%                      & $(77.6^{+3.4}_{-6.7})$\% \\ 
\hline 
\end{tabular}
\end{table} 

The Monte Carlo data samples are not purely composed of single-site or
multi-site events. The $DEP$ samples are dominated by single-site
events, the $\Gamma_{1}$ and $\Gamma_{2}$ have large fractions of
multi-site events. Events in the $DEP$ samples are referred to as {\it
electron-like} while events in the $\Gamma_{1}$ and $\Gamma_{2}$
samples are referred to as {\it photon-like} in the following. Note,
that these two labels do not describe an intrinsic property of an
event (such as the range of energy deposition), but they are used to
emphasize the different probabilities of the event being single-site
or multi-site.


\section{Analysis methods}
\label{section:methods}

Three analysis methods were tested. Half of the $DEP$ and $\Gamma_{1}$
data samples were used to train the methods. The other half of the
samples, together with the $\Gamma_{2}$ and $ROI$ samples, were used
to test the analysis methods. The $DEP$ and $\Gamma_{1}$ samples were
selected for training in order to avoid biases due to the difference
in energy of events in the two samples. For the same reason the
maximum of each pulse shape was normalized to unity for each event. \\

The analyses were applied to the core and segment data samples in
order to study the effect of pulse shape analysis before and after the
application of a single segment requirement. In the former case, only
the core pulse shape was used. In the latter case, the core pulse
shape was used and, optionally, the segment~$S$ pulse shape in
addition.

\subsection{Likelihood discriminant method} 

\begin{figure}[ht!]
\center
\mbox{\epsfig{file=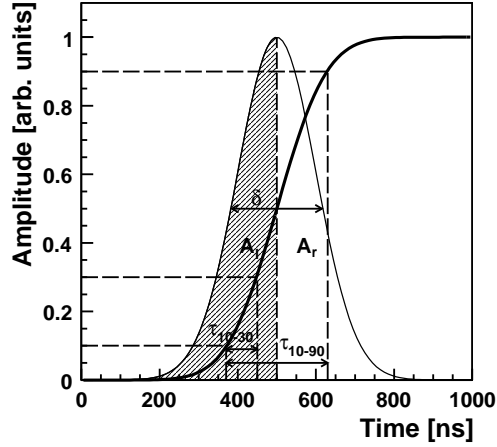,width=0.45\textwidth}}
\caption{Ideal pulse shape: the integrated charge (thick line) and the
current (thin line). Indicated are the quantities $\tau_{10-30}$,
$\tau_{10-90}$, $\delta$, $A_{l}$ and $A_{r}$ (see text).
\label{fig:quantities}}
\end{figure}

Four quantities are calculated for each pulse shape. These quantities
provided separation power in previous
studies~\cite{Aalseth:2000hy,Elliott:2005at}. Interpolation algorithms
were applied to the pulse shapes to obtain continuous
distributions. Figure~\ref{fig:quantities} shows an ideal pulse and
the quantities calculated are indicated. All quantities are given
subscripts $C$ and $S$ for the core and segment pulse shapes,
respectively.

\enlargethispage{0.5 cm}

\begin{itemize}
\item Risetime $\tau_{10-30}$, defined as the difference between the 
times the integrated charge amplitude has reached 10\% and 30\% of its
maximal amplitude;
\item risetime $\tau_{10-90}$, defined as the  difference between the 
times the integrated charge amplitude has reached 10\% and 90\% of its
maximal amplitude;
\item left-right asymmetry $\zeta$, defined as the asymmetry of the 
area below the left and the right half of the current pulse, $A_{l}$
and $A_{r}$, measured from the maximum\footnote{The definition differs
from the one given in~\cite{Aalseth:2000hy,Elliott:2005at}.}, $\zeta =
\frac{A_{l}-A_{r}}{A_{l}+A_{r}}$;
\item current pulse width $\delta$, defined as the full width at half maximum 
of the current pulse.
\end{itemize}

\begin{figure}[ht!]
\center
\begin{tabular}{cc}
\begin{minipage}[ht!]{0.45\textwidth}
\mbox{\epsfig{file=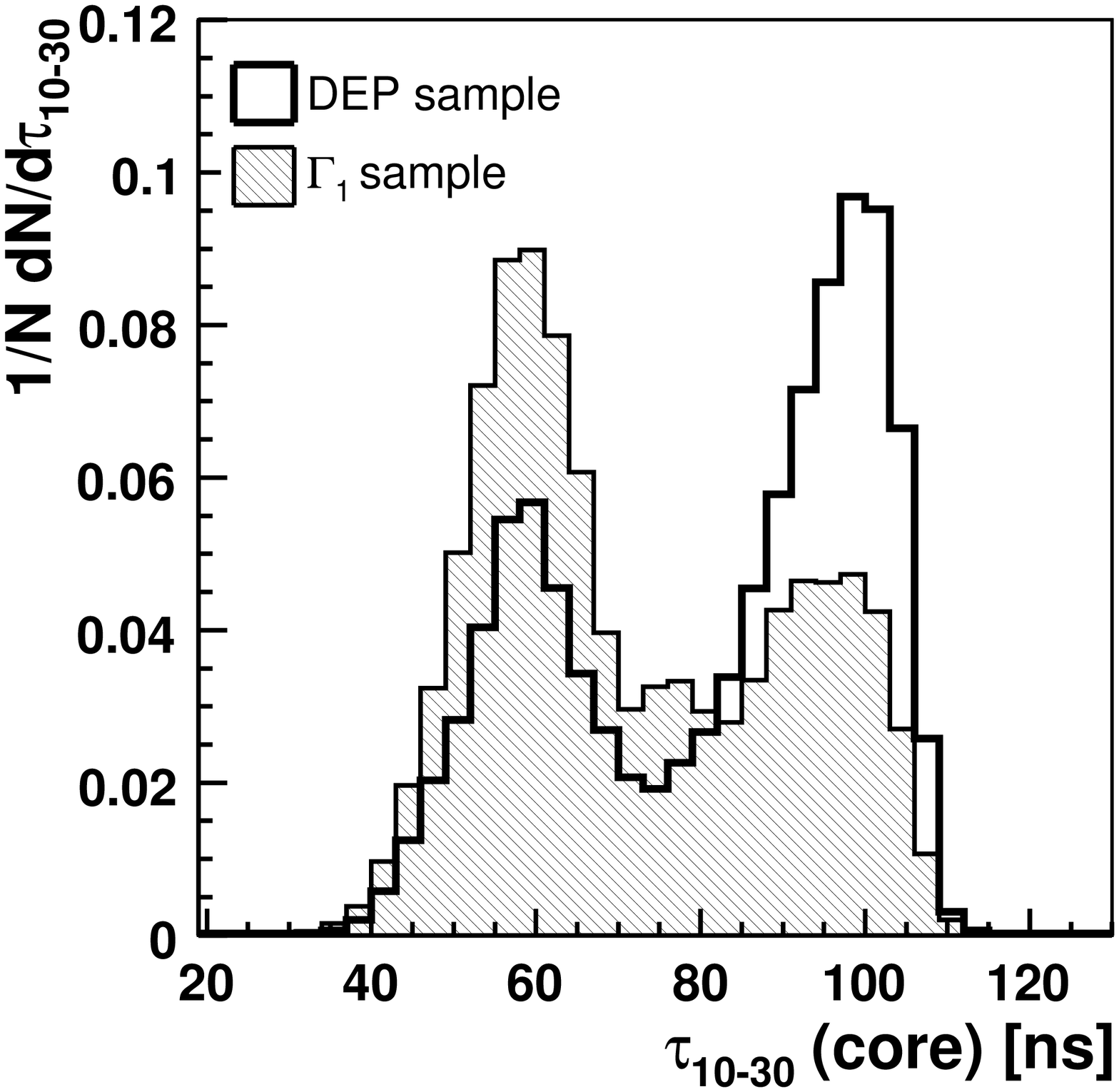,width=\textwidth}}
\end{minipage}
&
\begin{minipage}[ht!]{0.45\textwidth}
\mbox{\epsfig{file=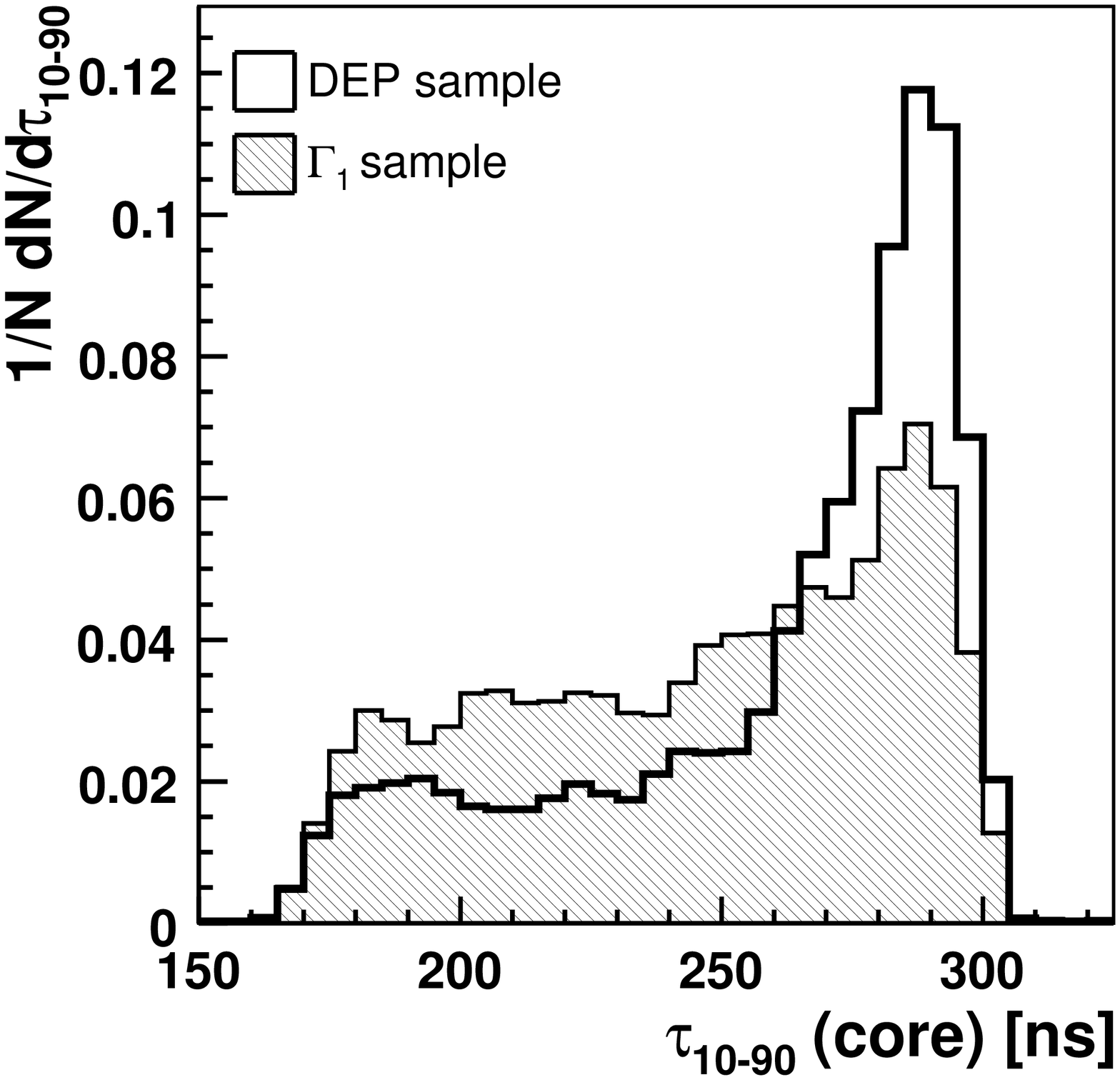,width=\textwidth}}
\end{minipage} \\
\begin{minipage}[ht!]{0.45\textwidth}
\mbox{\epsfig{file=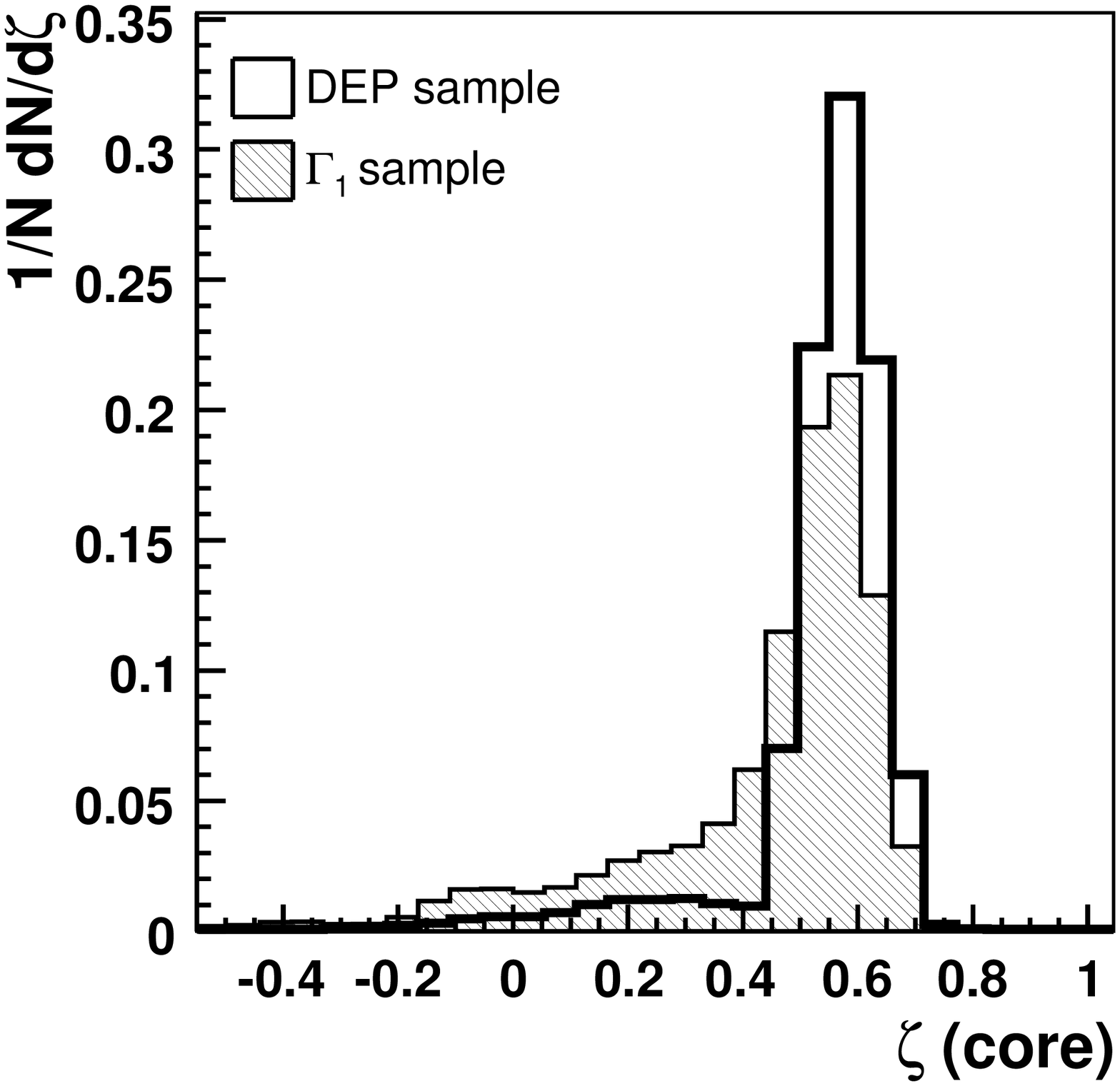,width=\textwidth}}
\end{minipage}
&
\begin{minipage}[ht!]{0.45\textwidth}
\mbox{\epsfig{file=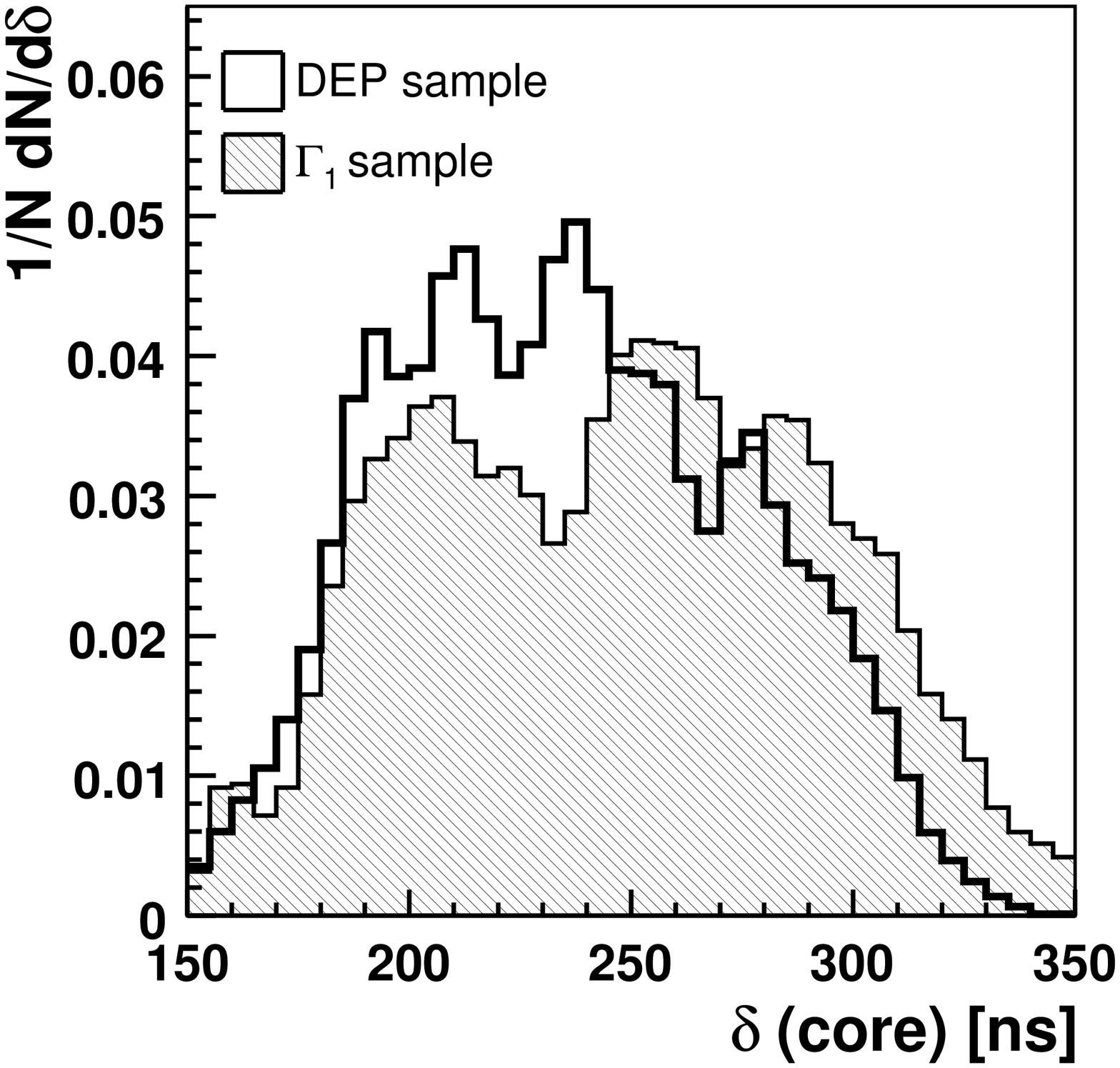,width=\textwidth}}
\end{minipage} 
\end{tabular}
\caption{Quantities calculated from the core pulseshapes in the $DEP$
(open histogram) and $\Gamma_{1}$ (hatched histogram) segment data
samples. Top left: risetime $\tau_{10-30}$, top right: risetime
$\tau_{10-90}$, bottom left: left-right asymmetry $\zeta$, bottom
right: current pulse width $\delta$.
\label{fig:quantities_distributions}}
\end{figure}

The variables are histogrammed for both training samples and their
integrals are normalized to unity.  As an example,
Figure~\ref{fig:quantities_distributions} shows the normalized
distributions of the four quantities calculated from the core pulse
shape in the two segment data samples. The average risetime of pulses
in the $DEP$ sample is larger than that of pulses in the $\Gamma_{1}$
sample~\footnote{This behavior was also found in a simple calculation
of pulse shapes assuming a perfect crystal and not taking into account
any effects from the electronics.}. The relative frequencies are used
to define discriminants, given that the event is electron-like ($DEP$
sample) or photon-like ($\Gamma_{1}$ sample). The respective overall
discriminants, $p_{e^{-}}$ and $p_{\gamma}$, are calculated by
multiplying the individual discriminants:

\begin{eqnarray}
p_{e^{-}}^{k} & = & p(\tau_{\mathrm{10-30},k}|e^{-}) \cdot p(\tau_{\mathrm{10-90},k}|e^{-}) \cdot 
                    p(\zeta_{k}|e^{-}) \cdot p(\delta_{k}|e^{-}) \ , \\ 
&& \nonumber \\ 
p_{\gamma}^{k} & = & p(\tau_{\mathrm{10-30},k}|\gamma) \cdot p(\tau_{\mathrm{10-90},k}|\gamma) \cdot 
                     p(\zeta_{k}|\gamma) \cdot p(\delta_{k}|\gamma) \ , 
\end{eqnarray}
with $k=C$~or~$S$ for the core and segment pulses, respectively. Note
that no correlations among these quantities are taken into account.

Likelihood discriminants (LHD) are constructed from $p_{e^{-}}$ and
$p_{\gamma}$ for each individual event:

\begin{eqnarray} 
D^{C}   & = & \frac{p_{e^{-}}^{C}}{p_{e^{-}}^{C} + p_{\gamma}^{C}} \ , \\ 
D^{C+S} & = & \frac{p_{e^{-}}^{C} \cdot p_{e^{-}}^{S}}{p_{e^{-}}^{C} \cdot p_{e^{-}}^{S}+ p_{\gamma}^{C} \cdot p_{\gamma}^{S}} \ , 
\end{eqnarray}
\noindent 
where $D^{C}$ uses information from the core electrode only and
$D^{C+S}$ uses information from the core and segment electrodes. $D$
varies between~0 and~1 by construction. $D$ peaks at~1 for
electron-like events; for photon-like events $D$ peaks at 0. Events
are identified as electron-like for $D>\overline{D}$ and as
photon-like for $D<\overline{D}$, where $\overline{D}$ is a chosen
parameter.

\subsection{Library method} 

The training $DEP$ samples are interpreted as libraries of
electron-like reference pulses. An average $\chi^{2}$ with respect to
all reference pulses is calculated for each pulse shape in the test
samples. For the $k$th reference pulse and the $l$th pulse shape under
study the average $\chi^{2}$ is defined as
\begin{equation}
\chi_{k,l}^{2} = \frac{1}{N}\sum_{i = 1}^{N} \frac{(x_{k, i} - x_{l,i})^{2}}{\sigma^{2}} \ ,
\end{equation} 
where $N$ is the number of bins of the pulse shapes and
$x_{k,i}$ and $x_{l,i}$ are the pulse heights in bin $i$ of
the $k$th reference pulse and the $l$th pulse under
study. $\sigma^{2}$ is defined as
\begin{equation}
\sigma^{2} = \sigma_{\mathrm{k}}^{2} + \sigma_{\mathrm{l}}^{2},
\end{equation}
where $\sigma_{\mathrm{k}}$ and $\sigma_{\mathrm{l}}$ are
the noise amplitudes of the reference pulse shape and the pulse shape
under study. The noise amplitude is the RMS of the baseline measured
during the one $\mu$s before the onset of the pulse. \\

The minimum $\chi^{2}$ is selected with respect to the reference
pulses and denoted
$\chi^{2}_\mathrm{min}=\chi^{2}_{k_{\mathrm{min}},l}$ for each pulse
shape in the test sample. Ideally, the minimum $\chi^{2}$ for
electron-like events should be smaller than that of photon-like
events. Events are identified as electron-like for
$\chi^{2}_{\mathrm{min}} <
\overline{\chi^{2}}$ and as photon-like for $\chi^{2}_{\mathrm{min}} >
\overline{\chi^{2}}$, where $\overline{\chi^{2}}$ is a chosen 
parameter.

\subsection{Neural network method} 

Artificial neural networks (ANNs) are used to separate electron-like
from photon-like events. Input neurons are fed with samples of the
normalized pulse shape, starting from the time when the amplitude has
reached 10\%. 40 consecutive samples per pulse shape are used. The ANN
consists of 40 input neurons, 40 hidden neurons and one output neuron
for the core data samples. An additional 40 input neurons are used
optionally for the segment data samples. \\

The ANNs are trained by feeding them with pulse shapes from the two
training samples and simultaneously providing the information which of
the samples each pulse belongs to (0:~$DEP$ sample, 1:~$\Gamma_{1}$
sample). The ANNs adjust the internal neurons iteratively using the
Broyden, Fletcher, Goldfarb, Shanno (BFGS) learning
method~\cite{BFGS}. Each ANN is trained in about
$1000$~iterations. The output quantity, $NN$, is on average larger
for photon-like events than for electron-like events. Events are
identified as electron-like for $NN < \overline{NN}$ and as
photon-like for $NN > \overline{NN}$, where $\overline{NN}$ is a
chosen parameter.

\pagebreak 


\section{Results} 
\label{section:results} 

\begin{figure}[ht!]
\center
\begin{tabular}{ccc}
\begin{minipage}[ht!]{0.30\textwidth}
\mbox{\epsfig{file=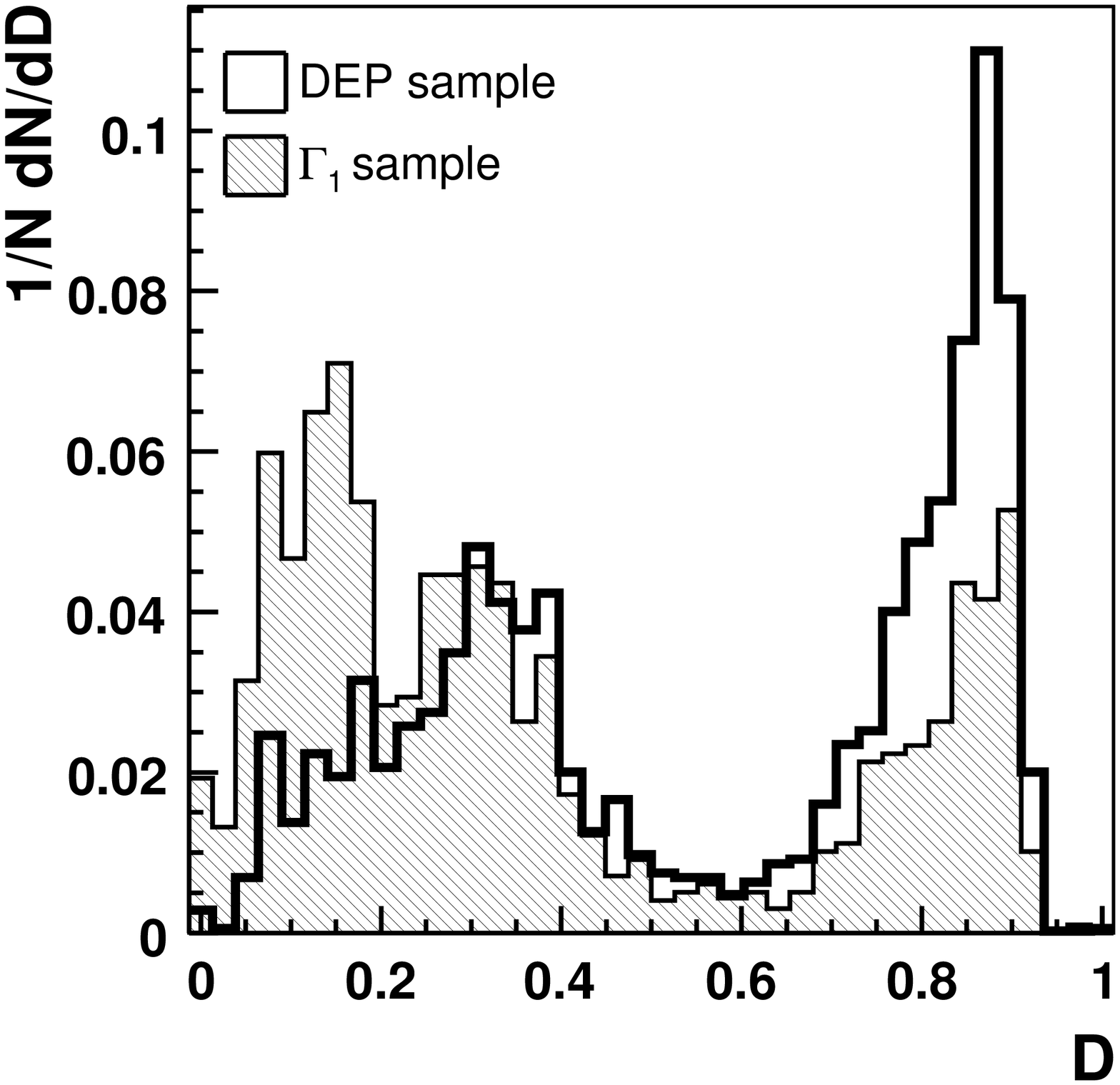,width=\textwidth}}
\end{minipage}
&
\begin{minipage}[ht!]{0.30\textwidth}
\mbox{\epsfig{file=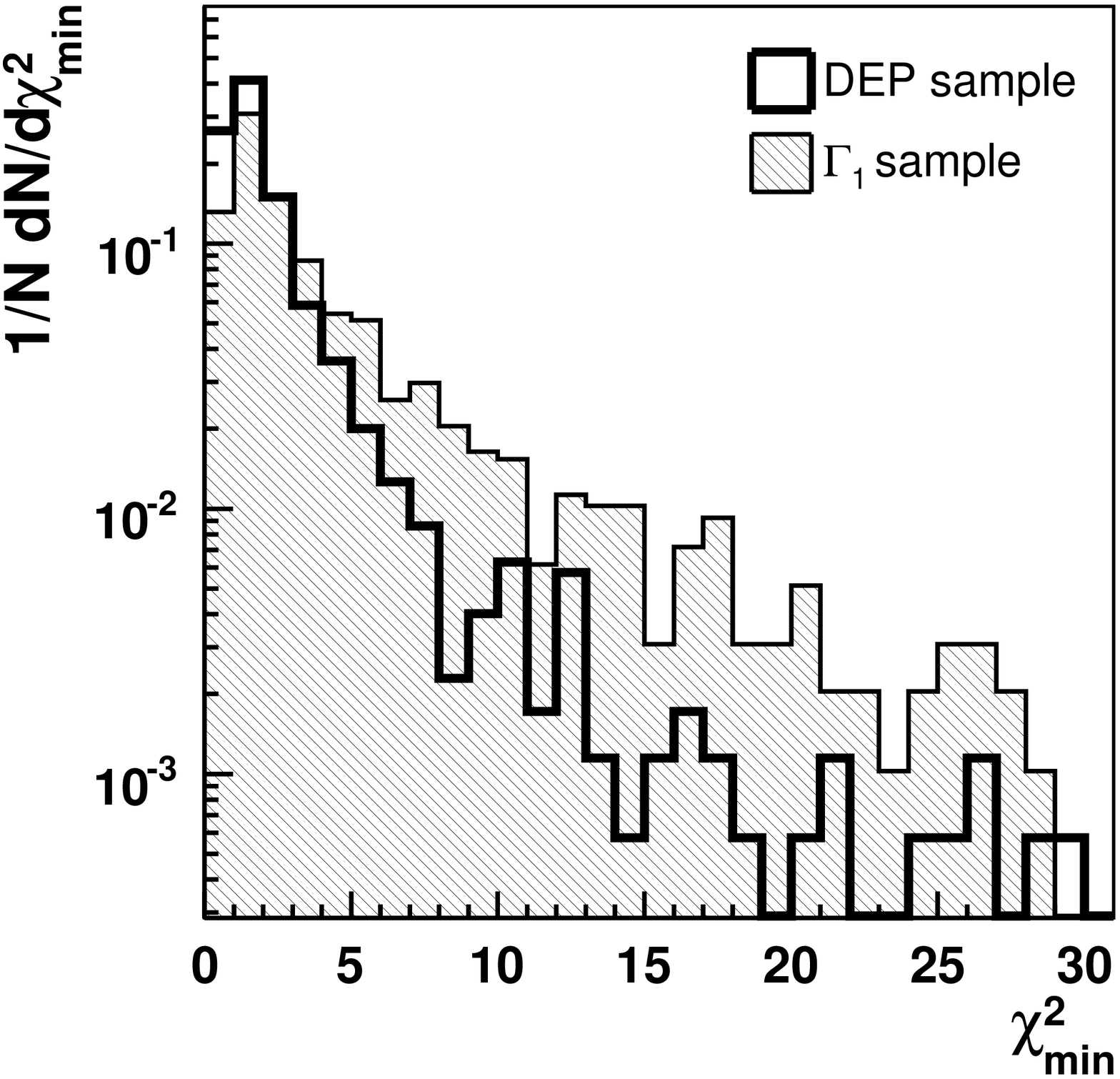,width=\textwidth}}
\end{minipage} 
&
\begin{minipage}[ht!]{0.30\textwidth}
\mbox{\epsfig{file=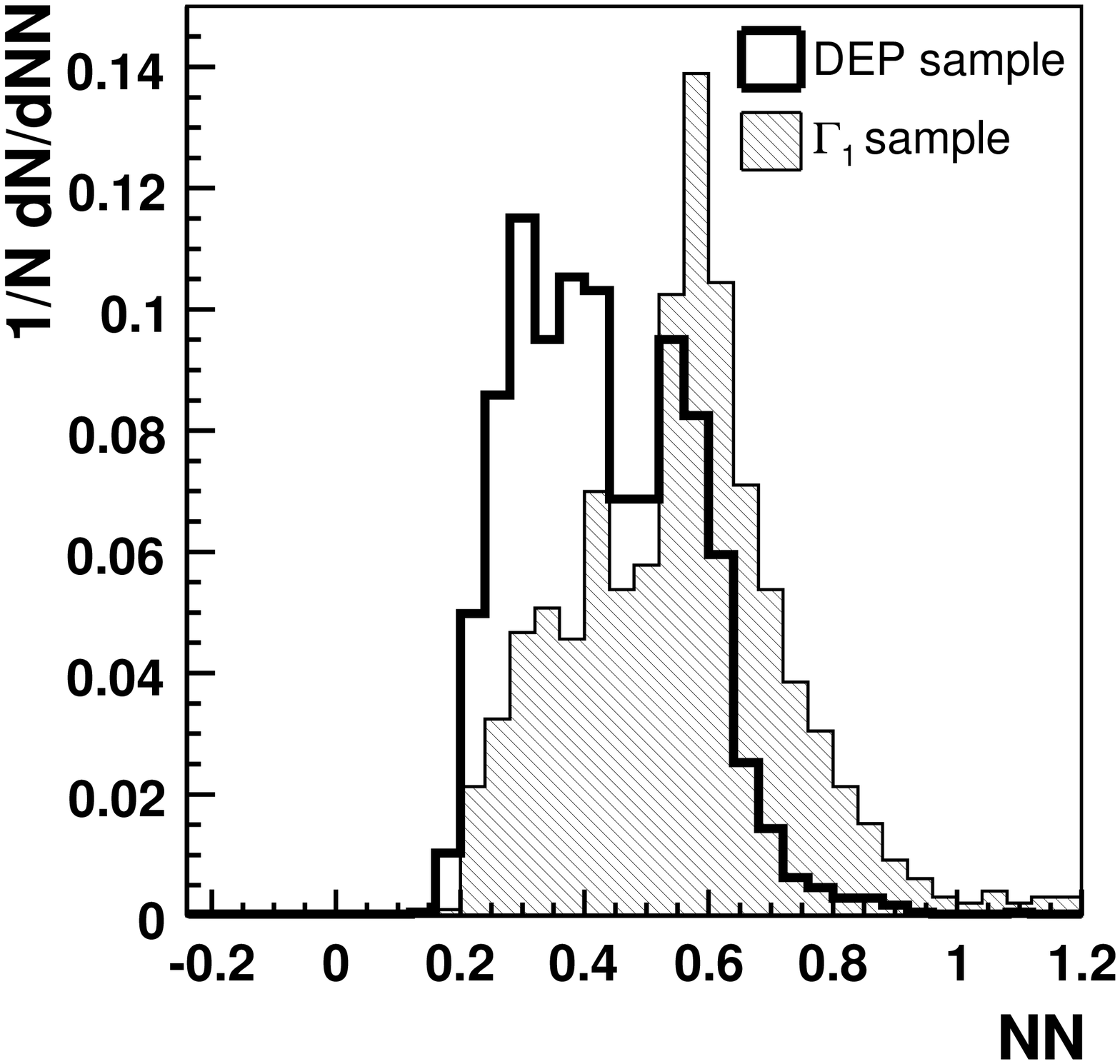,width=\textwidth}}
\end{minipage} \\
\end{tabular}
\caption{Output distributions for the two segment training data
samples $DEP$ (open histograms) and $\Gamma_{1}$ (hatched histograms)
for the likelihood method (left), the library method (middle) and the
neural network (right). The segment pulse shapes were not taken into
account in these examples.
\label{fig:PSA_output}}
\end{figure}

The three analysis methods are applied to the data samples defined in
Section~\ref{subsection:selection}. The likelihood discriminant and
neural network analysis are performed on the segment data samples (a)
with information from the core electrode only and (b) with information
from the core and the segment~$S$ electrode. As an example,
Figure~\ref{fig:PSA_output} shows the output distributions for the two
segment training data samples $DEP$ and $\Gamma_{1}$ for the
likelihood method (left), the library method (middle) and the neural
network (right). The segment pulse shapes have not been taken into
account for these examples.

The results of the analysis are interpreted in the following. First,
it is shown that the electron-like and photon-like event samples can
be distinguished. In a second step, the results are interpreted to
distinguish between single-site and multi-site events. The estimate of
the power of such a distinction requires the knowledge of the fraction
of single-site and multi-site events in the data samples. That
information is taken from the Monte Carlo simulation presented in
Section~\ref{section:simulation} based on the parameter $R_{90}$.

\subsection{Distinction between electron-like and photon-like event samples} 
\label{subsection:electronlike} 

\enlargethispage{0.5 cm} 

The power to distinguish between electron-like and photon-like event
samples is estimated. The events in the $DEP$ sample are assumed to
give the same output in the analyses as events from neutrinoless
double beta-decay. The cut values are chosen to keep 90\% of the
events in the $DEP$ training samples for the three analysis methods
and thus a high detection efficiency. The fraction of events in each
test data sample identified as electron-like are summarized in
Table~\ref{table:fractions}. The uncertainties are estimated from the
deviation from 90\% of the fraction of events identified as
electron-like in the $DEP$ test data samples and found to be about
2\%. Note that no deviation is found in case of the library method
since the $DEP$ training data sample is used as a reference library.

\begin{table}[ht!]
\center
\caption{Fraction of events in the test data samples identified as 
electron-like for the three analyses. The uncertainties are estimated
to be about 2\%. 
\label{table:fractions}}
\center
\begin{tabular}{lcccc}
\\
\hline
Data samples                    & $DEP$             & $\Gamma_{1}$    & $\Gamma_{2}$   & $ROI$ \\  
                                & ($1593$~keV)  & ($1620$~keV)  & ($2615$~keV) & (2039~keV) \\  
\hline 
Likelihood method & & & & \\ 
\hline 
Core samples                    & 89.3\% & 76.5\% & 75.4\% & 83.4\% \\ 
Segm. samples, core only        & 89.3\% & 67.1\% & 64.1\% & 84.8\% \\ 
Segm. samples, core \& segm. & 88.0\% & 66.7\% & 61.1\% & 83.4\% \\ 
\hline					 
Library method & & & & \\ 		 
\hline 					 
Core samples                    & 90.0\% & 86.9\% & 85.8\% & 86.7\% \\ 
Segm. samples, core only        & 90.0\% & 68.4\% & 56.4\% & 83.1\% \\ 
\hline					 
Neural network method & & & & \\ 	 
\hline 					 
Core samples                    & 90.4\% & 65.8\% & 63.2\% & 79.9\% \\ 
Segm. samples, core only        & 89.3\% & 54.1\% & 44.3\% & 80.8\% \\ 
Segm. samples, core \& segm. & 89.3\% & 56.1\% & 49.9\% & 79.6\% \\ 
\hline
\end{tabular}
\end{table}

The fraction of events identified as electron-like is significantly
lower than 90\% in the $\Gamma_{1}$, $\Gamma_{2}$ and $ROI$
samples. The fraction in the $\Gamma_{1}$ sample is found to be larger
than that in the $\Gamma_{2}$ sample with each method. This is
expected, as the mean free path of photons increases with the photon
energy. \\

The fraction of events identified as electron-like in the $\Gamma_{1}$
and $\Gamma_{2}$ segment data samples (using the core pulse shape
only) is found to be lower than that in the core data samples with all
three methods. The additional usage of the segment pulse shape in the
analyses reduces the fraction by maximally 3\%; in case of the neural
network it even increases the fraction by maximally 5\%. This
demonstrates that the additional information is highly correlated with
the existing information and only marginally contributes to the
analysis. \\

The neural network shows the best performance. This is expected, since
the ANN uses the largest fraction of information and also takes
correlations between input variables into account. \\

\subsection{Selection of single-site events and discrimination against multi-site events} 

As demonstrated in Table~\ref{table:fraction}, neither the $DEP$ nor
the $\Gamma_{1}$, $\Gamma_{2}$ and $ROI$ samples are solely composed
of single-site or multi-site events. The probability to correctly
identify single-site and multi-site events as such, $\epsilon$ and
$\eta$, can be deduced from the fraction of single-site and multi-site
events in each sample (estimated from Monte Carlo) and the output of
the analyses, $D$, $\chi^{2}_\mathrm{min}$, $NN$:

\begin{eqnarray}
\epsilon & = & \frac{ N_{id}^{SSE}/N_{true}^{MSE} - M_{id}^{SSE}/M_{true}^{MSE} }{ N_{true}^{SSE}/N_{true}^{MSE} - M_{true}^{SSE}/M_{true}^{MSE} } \ , \label{eqn:epsilon} \\ 
&& \nonumber \\ 
\eta     & = & \frac{ N_{id}^{MSE}/N_{true}^{SSE} - M_{id}^{MSE}/M_{true}^{SSE} }{ N_{true}^{MSE}/N_{true}^{SSE} - M_{true}^{MSE}/M_{true}^{SSE} } \ , \label{eqn:eta} 
\end{eqnarray} 

\noindent 
where $N_{id}^{SSE}$ and $N_{id}^{MSE}$ are the number of events in
the $DEP$ sample identified as single-site and multi-site events,
respectively. The numbers depend on the cut value chosen for each
analysis. $N_{true}^{SSE}$ and $N_{true}^{MSE}$ are the true number of
single-site and multi-site events in the same sample and are estimated
from the Monte Carlo simulation discussed in
Section~\ref{section:simulation}. $M_{id}^{SSE}$ and $M_{id}^{MSE}$
are the number of events in the $\Gamma_{1}$ sample identified as
single-site and multi-site events, respectively. $M_{true}^{SSE}$ and
$M_{true}^{MSE}$ are the true number of single-site and multi-site
events in the same sample. The probabilities $\epsilon$ and $\eta$ are
assumed to be the same for all samples. This assumption is reasonable
for the $DEP$ and $\Gamma_{1}$ samples as the average energies are
very close. \\

The cut values for the three analysis methods are chosen to maximize
the figure of merit, the identification efficiency
$\sqrt{\epsilon\cdot\eta}$. Note, that these cut values differ from
those used in Section~\ref{subsection:electronlike}. The probabilities
obtained from the data samples using Equations~\ref{eqn:epsilon}
and~\ref{eqn:eta} are listed in Table~\ref{table:efficiencies}.

\begin{table}[ht!]
\caption{Probabilities $\epsilon$ and $\eta$ obtained for all three 
analysis methods. The errors are introduced by the choice of
$\overline{R}$ determining the fraction of single-site and multi-site
events.
\label{table:efficiencies}}
\center
\begin{tabular}{lccc}
\\
\hline
Analysis                        & $\epsilon$     & $\eta$          & $\sqrt{\epsilon\cdot\eta}$ \\ 
\hline
Likelihood method & & & \\ 
\hline 
Core samples                    & ($74.8^{+1.8}_{-0.3}$)\% & ($\phantom{0}84.7^{+\phantom{0}3.4}_{-\phantom{0}2.4}$)\%  & ($79.6^{+1.4}_{-0.2}$)\% \\ 
Segm. samples, core only        & ($84.3^{+1.8}_{-0.2}$)\% & ($\phantom{0}97.7^{+10.4}_{-\phantom{0}5.9}$)\%  & ($90.8^{+4.8}_{-1.9}$)\% \\ 
Segm. samples, core \& segm. & ($83.9^{+1.7}_{-0.1}$)\% & ($\phantom{0}94.0^{+\phantom{0}9.9}_{-\phantom{0}5.6}$)\%  & ($88.8^{+4.6}_{-1.8}$)\% \\ 
\hline 
Library method & & & \\ 
\hline 
Core samples                    & ($68.7^{+\phantom{0}0.8}_{-\phantom{0}0.1}$)\%  & ($56.1^{+\phantom{0}1.4}_{-\phantom{0}1.0}$)\%  & ($62.1^{+0.7}_{-0.2}$)\% \\ 
Segm. samples, core only        & ($90.9^{+\phantom{0}0.1}_{-13.4}$)\% & ($80.4^{+10.1}_{-\phantom{0}9.1}$)\%  & ($85.6^{+4.8}_{-1.7}$)\% \\ 
\hline 
Neural network method & & & \\ 
\hline 
Core samples                    & ($85.6^{+2.4}_{-0.4}$)\% & ($\phantom{0}91.0^{+\phantom{0}4.3}_{-\phantom{0}0.3}$)\%  & ($\phantom{0}88.3^{+1.9}_{-0.3}$)\% \\ 
Segm. samples, core only        & ($96.4^{+2.5}_{-0.2}$)\% & ($121.6^{+15.0}_{-\phantom{0}8.5}$)\%  & ($108.3^{+6.6}_{-2.5}$)\% \\ 
Segm. samples, core \& segm. & ($90.6^{+2.3}_{-0.2}$)\% & ($115.4^{+13.4}_{-\phantom{0}7.7}$)\%  & ($102.3^{+5.9}_{-2.2}$)\% \\ 
\hline
\end{tabular}
\end{table}

The likelihood and library methods work better on events with only one
segment hit. The additional usage of the segment pulse shape in the
likelihood method does not improve the analysis results. \\

\begin{figure}[ht!]
\center
\epsfig{file=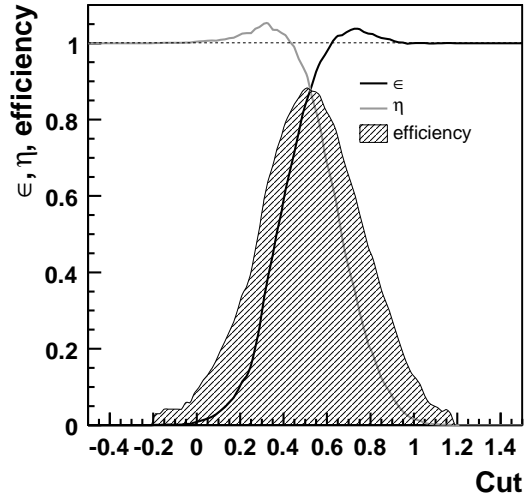,width=0.5\textwidth}
\caption{Probabilities to correctly identify single-site, $\epsilon$,
and multi-site events, $\eta$, and the efficiency,
$\sqrt{\epsilon\cdot\eta}$, for the neural network analysis of the
core data samples. Probabilities above one are caused by uncertainties
in the extraction process.
\label{fig:PSA_efficiency}}
\end{figure}

The analysis of the neural network output yields probabilities larger
than one for the segment data samples. The calculation of $\epsilon$
and $\eta$ depends on the real fraction of single-site and multi-site
events and is therefore model dependent. The current model assumes the
fraction of single-site and multi-site events to be completely
reflected by the parameter $R_{90}$. The validity of the assumed model
is limited and the extraction of the probabilities $\epsilon$ and
$\eta$ carries systematic uncertainties. The results should be taken
with care. The efficiencies do not exceed unity for the chosen cut
parameter for the core data samples. Figure~\ref{fig:PSA_efficiency}
shows $\epsilon$ and $\eta$ together with the identification
efficiency as a function of the neural network cut parameter for the
core data samples.

\subsection{Application to the $^{228}$Th data set} 
\label{subsection:application} 

\begin{figure}[ht!]
\center
\begin{tabular}{cc}
\begin{minipage}[ht!]{0.45\textwidth}
\mbox{\epsfig{file=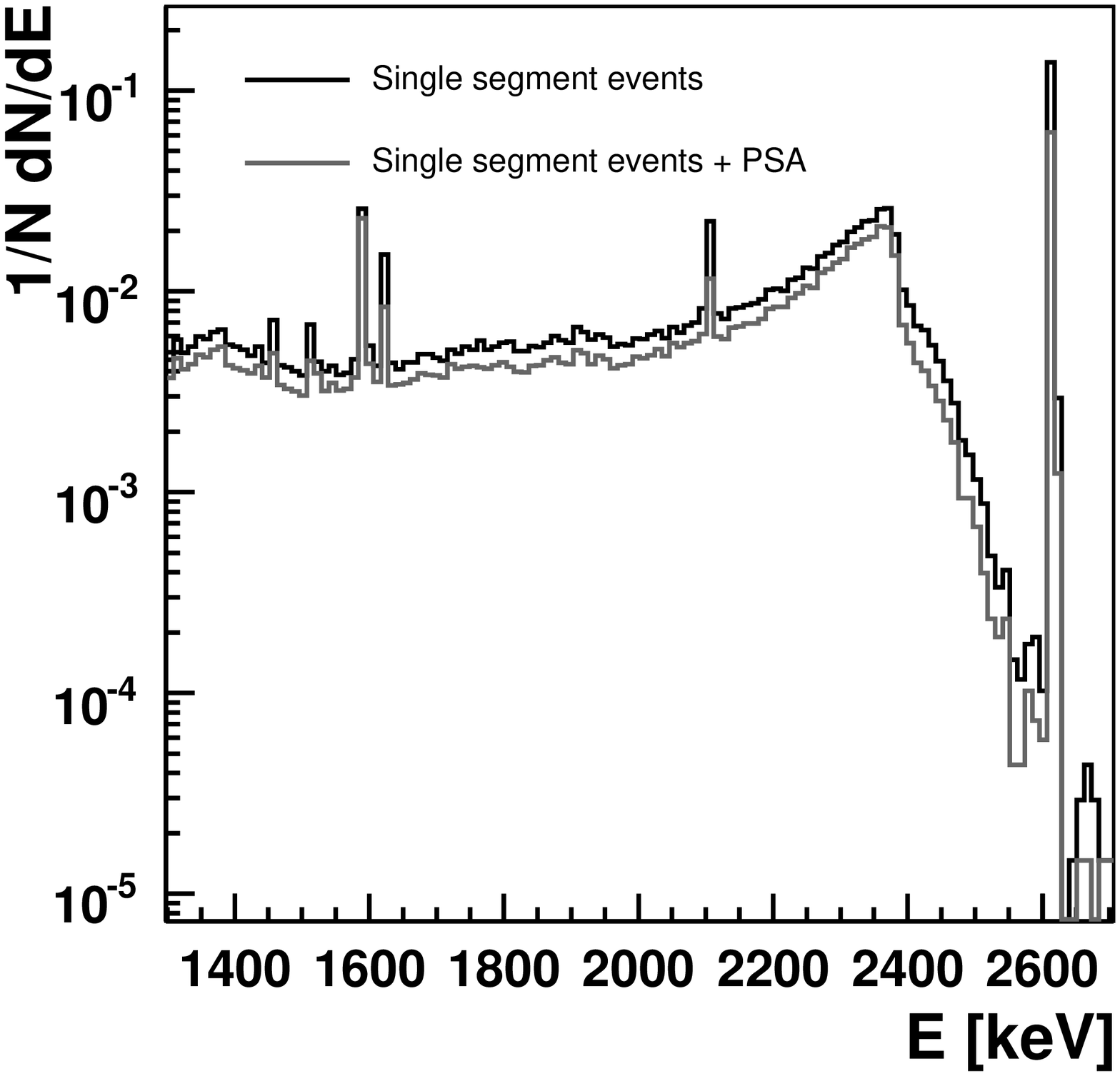,width=\textwidth}}
\end{minipage}
&
\begin{minipage}[ht!]{0.45\textwidth}
\mbox{\epsfig{file=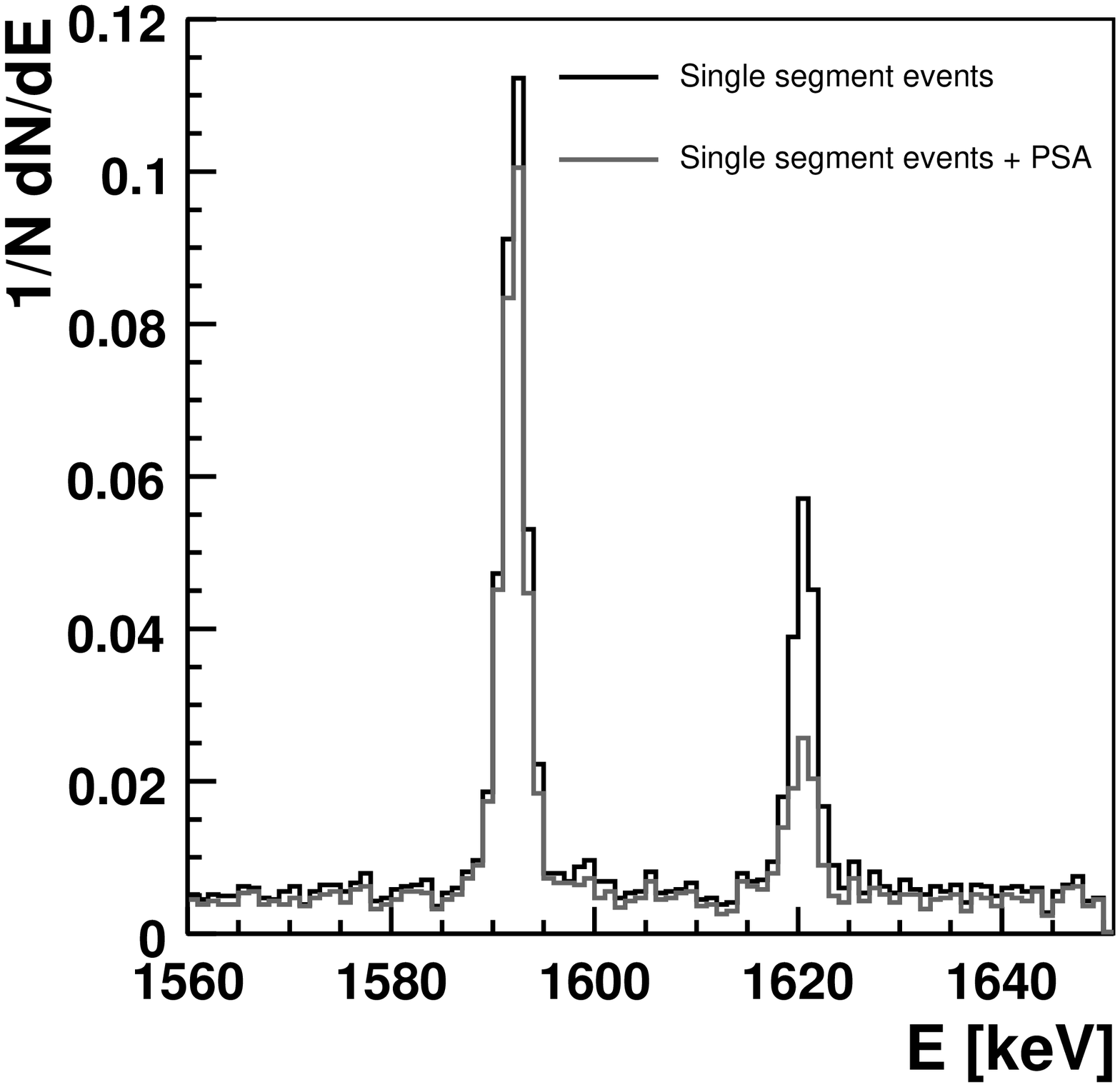,width=\textwidth}}
\end{minipage} \\
\end{tabular}
\caption{Spectrum of a $^{228}$Th source as seen by the core
electrode. The black line corresponds to all events with only
segment~$S$ hit, the gray line represents events with only segment~$S$
hit and pulse shape analysis, using the ANN, applied. Only the pulse
shape of the core was used and the cut parameter was chosen to keep
90\% of the $DEP$ events. Left: Spectrum from 1.3~MeV to
2.7~MeV. Right: Close-up of the region from 1560~keV to 1650~keV. For
a discussion see text.
\label{fig:spectrum}}
\end{figure}

Figure~\ref{fig:spectrum} (left) shows the energy spectrum resulting
from a $^{228}$Th source in the region from 1.3~MeV to 2.7~MeV as seen
by the core electrode. The black line corresponds to all events with
only segment~$S$ hit, the gray line represents events with only
segment~$S$ hit and pulse shape analysis, using the ANN, applied. Only
the pulse shape of the core was used and the cut parameter was chosen
to keep 90\% of the events in the $DEP$ training data sample. \\

The gray spectrum is suppressed with respect to the black
spectrum. The suppression ranges up to a factor of about two at the
photon peaks. The suppression is weak in the double escape
peak. Figure~\ref{fig:spectrum} (right) shows a close-up of the
spectrum in the region from 1560~keV to 1650~keV. The application of
the pulse shape analysis removes photon induced events (1620~keV
photon line from the decay of $^{212}$Bi) but keeps most of the
electron induced events (double escape peak of the $2\ 615$~keV
$^{208}$Tl photon at $1593$~keV). Pulse shape analysis is thus
suitable to confirm the signal process. \\

%

\pagebreak 

\section{Conclusions and outlook}
\label{section:conclusions}

Three methods using pulse shapes were introduced to distinguish
electrons from multiply scattered photons. They were applied on data
collected with a {\sc GERDA} prototype detector.  Single-site
dominated samples were distinguished from multi-site dominated
samples. The probability to correctly identify single-site and
multi-site events was estimated based on Monte Carlo calculations. \\

All three methods were trained with double escape events and events
from a nearby photon peak. The former events are expected to be
similar to the expected $0\nu\beta\beta$-events. \\

The methods are based on information from the core electrode and may
include information from the segment electrode or not. The power to
identify photon induced events does not increase with the
straightforward inclusion of the pulse shape of the segment. \\

The performance of the three methods is slightly worse than what was
reported in~\cite{Elliott:2005at}. A reason for this is the purity of
the samples. Also, the spatial distribution of energy deposited inside
the detector is not homogeneous in the $DEP$ sample. Methods to select
cleaner and more homogeneous training samples are currently being
tested. \\

The artificial neural network shows a better performance than both the
likelihood discriminant and the library method. Photon peaks remaining
after a single segment cut are suppressed by a factor of about two at
energies around 1.5~MeV. At the same time 90\% of the events in the
single-site dominated sample are kept. This demonstrates that the
association of a particular peak with the signal process can be
substantiated by this kind of analysis. \\

The calculation of the efficiency to correctly identify single-site
and multi-site events is limited by the assumed model based on the
$R_{90}$ parameter. Further studies are required; in particular, a
simulation of the development of pulse shapes is important and is
currently under development. Studies using additional information from
neighboring segments to distinguish single-site from multi-site events
are also planned. In addition, an improved experimental setup is
planned. \\

The rejection of events in the $1620$~keV peak using segment
anti-coincidences as presented in~\cite{Abt:2007rg} is about a factor
of two better than the sole application of pulse shape analysis as
presented in this paper. Nevertheless, the application of pulse shape
analysis after a single segment cut can further reject events in this
peak by an additional factor of about two.


\section{Acknowledgements}

The authors would like to thank A.~Bettini, P.~Grabmayr, L.~Pandola
and B.~Schwingenheuer for their helpful comments and suggestions. The
authors would also like to thank the {\sc GERDA} and {\sc Majorana}
Monte~Carlo groups for their fruitful collaboration and cooperation on
the {\sc MaGe} project.


\addcontentsline{toc}{section}{Bibliography}
%


\end{document}